\documentclass[tightenlines,aps,nofootinbib]{revtex4}

\usepackage{psfig,axodraw,float}
\usepackage{graphics}

\def\as{\alpha_s}

\def\Q{{\mathcal{Q}}}
\def\D{{ D}^{\rm ini}}
\def\MSbar{\overline{\mathrm{MS}}}

\def\e{\epsilon}

\def\Li{{\rm Li}}

\begin{document}

\begin{flushright}
\vbox{
\begin{tabular}{l}
UH-511-1059-2004\\
\   \end{tabular} }
\end{flushright}

\vspace{0.6cm}

\title{
Perturbative Heavy Quark Fragmentation Function through ${\cal
O}(\alpha_s^2)$:\\
Gluon Initiated Contribution }

\author{Alexander Mitov $\!\!$\thanks{
e-mail:  amitov@phys.hawaii.edu}    }

\affiliation{Department of Physics and Astronomy,\\ University of Hawaii,\\
Honolulu, HI, 96822}

\begin{abstract}

\vspace{2mm}

We derive the gluon initiated contribution to the initial
condition for the perturbative fragmentation function of a heavy
quark through order ${\mathcal{O}}(\as^2)$ in the $\overline {\rm
MS}$ scheme. This result is needed for the resummation with
next-to-next-to-leading logarithmic accuracy of quasi-collinear
logarithms $\ln^k(m^2)$ in heavy quark differential distributions
by solving the complete DGLAP equation. Together with the
previously evaluated fermion initiated components, this result
completes the derivation of the initial condition for the
perturbative fragmentation function at next-to-next-to-leading
order.
\end{abstract}

\maketitle

\thispagestyle{empty}

\section{Introduction}

Production of heavy flavors (charm and bottom) in high energy
processes has become an increasingly important subject in the last
decade. In this type of  processes, not-completely inclusive
observables like energy or transverse momentum distributions of
heavy flavored hadrons are of special interest. The reason for
that lies in their dependence on both the short-distance
perturbative physics and the long-distance physics governing the
formation of the heavy flavored hadrons. A peculiar feature of
these observables is that they contain logarithmic sensitivity to
the mass of the heavy flavor to all orders in perturbation theory.
For that reason perturbative calculations of such cross-sections
are not reliable when the typical hard scale in the process is
much larger than the mass of the produced heavy quark.

A formalism for the study of differential distributions for heavy
quark production was proposed long ago by Mele and Nason
\cite{MN}. It is known as the Perturbative Fragmentation Function
(PFF) formalism and its idea is that in order to restore the
applicability of perturbative QCD to the type of observables
discussed above, one has to resum classes of large logarithms to
all orders in perturbation theory. The method is process
independent and is based on the DGLAP evolution equation
\cite{DGLAP}.

The main goal of this paper is to derive the gluon initiated
component of the initial condition for the perturbative
fragmentation function at order ${\mathcal{O}}(\as^2)$. Combined
with the fermion initiated contributions \cite{MM} at order
${\mathcal{O}}(\as^2)$, and with the three-loop time-like
splitting functions (when the latter become available), this
result can be used to perform resummations of the large collinear
logs with next-to-next-to-leading logarithmic (NNLL) accuracy.

In general, the importance of the gluon initiated contribution
increases in processes with larger separation between the hard
scale and the quark mass or in processes where gluon production is
not suppressed compared to the production of heavy quarks. This
expectation is supported from the previous applications of the PFF
formalism at the NLL accuracy level. The effect of the inclusion
of the gluon component is certainly important for hadron colliders
\cite{bquark, CG93}. The effect of the gluon component in $e^+e^-$
has been discussed, for example in \cite{CG97}. Therefore, at the
NNLL precision level, both the effect of the gluon initiated
component and the quark-gluon mixing will be important. In fact,
as we demonstrate in this paper, the property of the initial
condition for the PFF -- that its gluon initiated component at
order ${\mathcal{O}}(\as)$ vanishes for $\mu_0=m$, does not
persist at order ${\mathcal{O}}(\as^2)$.

This paper is organized as follows: in the next section we review
some general results and then describe the derivation of the gluon
initiated component. In section III we present our result and
discuss its properties. At the end we present our conclusions.

\section{Process-independent derivation of $\D_g$}

Let us consider heavy flavor production in a hard scattering
process. We are interested in the processes for which the
characteristic hard scale $Q$ is much larger than the mass $m$ of
the heavy quark. It follows from the QCD factorization theorems
\cite{ftheorem,Ellis,Collins,purplebook} that the non-perturbative
contributions to the energy distribution of the heavy-flavored
hadron produced in such reaction are contained in the so-called
non-perturbative fragmentation function. That function describes
the formation of the observed heavy flavored hadron from the heavy
quark $\Q$ at a scale set by the mass $m$, and is typically
extracted from $e^+e^-$ data. In this paper we will not be
concerned with the non-perturbative component of the fragmentation
function; details about its implementation can be found in
\cite{MN,CC,bquark,CG93,CG97,CM,a1,a2,CGardi,a3,a4}.

We will be only interested in the perturbative production of a
heavy quark with mass $m<<Q$. In such kinematics all power
corrections $\sim (m/Q)^p$ are neglected. This means that at large
scales the heavy quark behaves as massless and the role of its
mass is, essentially, to regulate the appearing collinear
divergences.

For the energy distribution of the heavy quark one writes:
\begin{equation}
{d\sigma_\Q \over dz}(z,Q,m) = \sum_a\int_z^1{dx\over x}
{d\hat{\sigma}_a \over
dx}(x,Q,\mu){D}_{a/\Q}\left(\frac{z}{x},\frac{\mu}{m} \right) +
{\cal O}\left({m\over Q}\right)^p. \label{fac}
\end{equation}
Here $m$ is the mass and $z=E_\Q/E_{\Q , max}$ the energy fraction
of the heavy quark produced in the reaction. The sum in
Eq.(\ref{fac}) runs over all partons -- quarks, antiquarks
(including the flavor $\Q$) and gluons -- that can be produced in
the hard process and $\mu$ is the factorization scale. The
function $d\hat\sigma_a/dx$, also known as coefficient function,
is the collinearly renormalized differential cross-section for
producing a massless parton $a$. The coefficient function is
related to the (bare) cross-section $\sigma_a (z,Q,\epsilon)$ for
producing a massless parton of type $a$ in the same reaction
through the following relation:
\begin{equation}
{d\sigma_a \over dz}(z,Q,\epsilon) = \sum_b {d\hat\sigma_b \over
dz}(z,Q,\mu)\otimes\Gamma_{ba}(z,\mu,\epsilon). \label{facN0}
\end{equation}
Eq.(\ref{facN0}) makes explicit the factorization of the collinear
singularities present in the massless cross-section
$d\sigma_a/dz(z,Q,\epsilon)$. All IR divergences are regularized
by working in $d=4-2\e$ dimensions. Throughout this paper we will
work in the ${\overline {\rm MS}}$ scheme where the collinear
counterterms $\Gamma_{ab}$ take the form:
\begin{widetext}
\begin{eqnarray}
&& \Gamma_{ba}= \delta_{ab}\delta(1-z) - \left({\as\over
2\pi}\right)\frac{P^{(0)}_{ab}(z)}{\epsilon} +\left({\as\over
2\pi}\right)^2 \left[ {1\over 2 \epsilon^2}\left(
P^{(0)}_{ac}\otimes P^{(0)}_{cb}(z) +\beta_0 P^{(0)}_{ab}(z)
\right) -{1\over 2\epsilon} P^{(1)}_{ab}(z)\right]. \label{Gamma}
\end{eqnarray}
\end{widetext}
Above, $\as = \as(\mu)$ stands for the strong coupling constant
renormalized in the $\MSbar$ scheme. Its relation to the bare
couplings can be found in \cite{MM}. The renormalized coupling
evolves with $n_f$ fermion flavors (including $\Q$), $\beta_0 =
(11C_A-4T_Rn_f)/6$ is the ${\cal O}(\alpha_s^2)$ coefficient of
the QCD $\beta$-function, $C_A=3,~T_R=1/2$ and $C_F=4/3$ are the
QCD color factors and $P_{ab}^{(0,1)}$ are the time-like splitting
functions \cite{furmanski}. Our notations for the splitting
functions follow Ref.\cite{purplebook}.

The functions ${D}_{a/\Q}(x,\mu/m)$  in Eq.(\ref{fac}) are the
perturbative fragmentation functions \cite{MN}. They are solutions
of the DGLAP evolution equation and satisfy an initial condition
given at some scale $\mu =\mu_0$:
\begin{equation}
{ D}_{a/\Q}\left ( z,\frac{\mu_0}{m} \right ) ={D}^{\rm ini}_{a}
\left ( z,\frac{\mu_0}{m} \right ). \label{D}
\end{equation}

Once the factorization scheme for the collinear singularities has
been chosen, the initial condition ${D}^{\rm ini}_{a}$ is an
unique, process independent function that has to be obtained from
an additional calculation. A prescription for the evaluation of
the initial condition can be extracted from the PFF method. The
key observation is that if the initial scale $\mu_0$ is chosen to
be of the order of the mass $m$, then the initial condition  can
not contain large logarithms and therefore can be derived from
fixed order perturbative calculations. In particular, ${D}^{\rm
ini}_{a}$ can be cast in the following form:
\begin{equation}
{D}^{\rm ini}_{a}\left (z,\frac{\mu_0}{m} \right ) = \sum_{n=0}
\left({\as(\mu_0)\over 2\pi}\right)^n d^{(n)}_{a} \left
(z,\frac{\mu_0}{m} \right ). \label{expansionD}
\end{equation}
Note that from the above equations and from the LO result
$d^{(0)}_{a}\sim\delta_{a\Q}$ it also follows that the knowledge
of ${D}^{\rm ini}$ to a given perturbative order allows one to
perform $\MSbar$ scheme subtraction (at the same order) in
calculations where the collinear divergences are regulated with
non-zero quark mass (see for example Ref.\cite{an} for more
details on that point).

To evaluate the initial condition one needs to combine
Eqs.(\ref{fac}) and (\ref{facN0}) and truncate Eq.(\ref{fac}) to a
fixed perturbative order by identifying the fragmentation function
with its initial condition. As a result, ${D}^{\rm ini}_{a}$ can
be expressed as a convolution of the following perturbative
factors: a known universal counterterm $\Gamma_{ab}$ and two
cross-sections $d\sigma_\Q$ and $d\hat\sigma_a$ that can be
separately evaluated in perturbation theory. Using results for
heavy and massless quark production in $e^+e^-$, the functions
$d^{(1)}_{a}$ were first obtained within this approach \cite{MN}.

Although this method works to any perturbative order, its
application beyond next-to-leading (NLO) order is impractical due
to the increased complexity in the evaluation of massive
cross-sections at higher orders in perturbation theory. A better,
process independent method for the calculation of the initial
condition $d^{(n)}_{a}$ was suggested in \cite{CC,KL} and further
developed in \cite{MM} where the functions $d^{(2)}_{a},~
a=\Q,\overline{\Q},q,\bar{q}$ were calculated. The method is based
on the observation that one needs to compute only a sufficient,
process independent part of the cross-sections for massive
(respectively massless) quark production. Moreover, this
calculation can be formulated in a process independent way. The
main purpose of this paper is to evaluate the function
$d^{(2)}_{g}$ using this approach and thus to complete the
derivation of the initial condition of the PFF through order
${\cal O}(\alpha_s^2)$. With this result and after the three-loop
time-like splitting functions become available, one will be able
to evaluate the perturbative fragmentation function with
next-to-next-to leading logarithmic accuracy. It follows from
Eq.(\ref{fac}) that the knowledge of the PFF with such accuracy
will make it possible to evaluate heavy quark spectra at NNLO in
any process and up to power corrections ${\cal O}(m/Q)$ from pure
massless calculations and resum large quasi-collinear logs
$\ln^k(m^2/Q^2)$ with NNLL accuracy to all orders in the strong
coupling constant.

Next, we describe the process independent derivation of the gluon
initiated component $d^{(2)}_{g}$. As was shown in \cite{MM}, the
initial condition for the perturbative fragmentation function can
be written as:
\begin{equation}
\D_{g}\left (z,{\mu_0 \over m} \right) = \sum_b
\Gamma_{gb}(z,\mu_0) \otimes \widetilde{D}_{b/\Q}\left (z,{\mu_0
\over m} \right), \label{DHresult}
\end{equation}
with $\Gamma_{ab}$ given in Eq.(\ref{Gamma}).

The function $\widetilde{D}_{b/\Q}$ can be interpreted as a bare
fragmentation function for production of a massive quark $\Q$ in
the decay of an (off-shell) parton of type $b$. That function
embodies all the effect of collinear radiation with $q_{\bot}\sim
m \leq \mu_0$ in that process. In general, the collinear limit is
defined as the kinematic limit where the relative transverse
momentum of two or more particles vanishes. When masses are
introduced, the collinear limit is \cite{CDT}:
\begin{equation}
q_{\bot} = \kappa q_{\bot} ,~ m = \kappa m ,~ \kappa \to 0.
\label{colllim}
\end{equation}

The following Sudakov parametrization is particularly convenient
for the discussion of the collinear limit in the process $g^*\to
\Q+X$ at order ${\cal O}(\alpha_s^2)$. Consider the one-to-three
collinear splitting of a gluon. We denote the four-momenta of the
produced massive quark and antiquark by $q_{1,2}$ while $q_3$
denotes the momentum of the radiated gluon. The momentum of the
decaying off-shell gluon is denoted as $\widehat{p}~$:
\begin{equation}
\widehat{p}=q_1+q_2+q_3. \label{phatsum}
\end{equation}
We parameterize the collinear direction by $p$ and introduce
another light-like vector $n$ as the complimentary light-cone
vector. We then write:
\begin{eqnarray}
q_i &=& z_ip + \beta_i {n\over (p n)} + q_{i,\bot} .
\label{sudakov}
\end{eqnarray}
The components $\beta_i$ are found from the on-shell conditions
$q_i^2=m_i^2$:
\begin{equation}
\beta_i = {m_i^2-q^2_{i,\bot}\over 2z_i}~~,~~~ i=1,2,3 ~,
\nonumber
\end{equation}
with $p^2=0$. We are interested in the symmetric case:
\begin{equation}
z_1+z_2+z_3 = 1~ ,~~ q_{1,\bot}+q_{2,\bot}+ q_{3,\bot} = 0 .
\label{constr}
\end{equation}
Using the above equations, we express the momentum $\widehat{p}$
through $p$ and $n$:
\begin{equation}
\widehat{p} = p + {1\over 2}\widehat{p}^2{n\over (pn)},
\label{phat}
\end{equation}
where:
\begin{eqnarray}
\widehat{p}^2= \frac{2(pq_2)+ 2(pq_3)-2(q_2q_3)  }{z}. \nonumber
\end{eqnarray}
We often write $z$ instead of $z_1$ to denote the measured energy
fraction of the heavy quark in the final sate.
Eqs.(\ref{sudakov},\ref{constr}) imply the energy conservation
condition:
\begin{equation}
z = 1 - {(nq_2)\over (pn)} -{(nq_3)\over (pn) }. \label{z}
\end{equation}
The kinematics of the one-to-two splitting can be obtained by
setting the momentum $q_3$ to zero in Eqs.(\ref{phatsum}-\ref{z}).

The explicit expression for the bare fragmentation function
appearing in Eq.(\ref{DHresult}) follows directly from the
factorization of phase-space and matrix elements in the collinear
kinematics. Consider a hard scattering process characterized by
some scale $Q \gg m$. Suppose that $n+2$ partons with momenta
$k_1,\dots,k_{n-1},q_1,q_2,q_3$ are produced. We consider momenta
$k_1,\dots,k_{n-1}$ as  non-exceptional, while momenta
$q_1,q_2,q_3$ are collinear, as described above. We denote the
phase-space of $n+2$ partons as ${\rm
dPS}^{(n+2)}(k_1,\dots,k_{n-1},q_1,q_2,q_3)$. In the limit when
$q_1,q_2$ and $q_3$ become collinear and are resulting from the
decay of an off-shell gluon,
$q_1+q_2+q_3=p+{\mathcal{O}}(q_\bot)$, the $(n+2)$-particle phase
space factorizes:
\begin{eqnarray}
{\rm dPS}^{(n+2)}(k_1,\dots,k_{n-1},q_1,q_2,q_3) = {\rm
dPS}^{(n)}(k_1,\dots,k_{n-1},p)~ {1\over z} [dq_2;m][dq_3;0].
\label{PSfact}
\end{eqnarray}
The factor $[dq;m_q]$ in Eq.(\ref{PSfact}) is the $d$-dimensional
one-particle phase space:
\begin{equation}
[dq;m_q] = {d^dq\over (2\pi)^{d-1}}\delta^+(q^2-m_q^2) .
\label{onePS}
\end{equation}

We next discuss the factorization of matrix elements in the
collinear kinematical configuration described above. As follows
from Eq.(\ref{fac}), we are not interested in power-suppressed
contributions to the cross-section. To identify the
logarithmically enhanced terms we apply well known power counting
arguments \cite{Ellis} in the collinear limit for the scattering
amplitudes: we rescale all momenta and masses according to
Eq.(\ref{colllim}), and then identify the leading contributions of
the amplitudes in the limit $\kappa \to 0$. Throughout the paper
we work in the light-cone
\footnote{For simplicity, the gauge fixing vector $n$ is chosen to
be the same vector that appears in the Sudakov parametrization
Eq.(\ref{sudakov}).}
gauge $n_\mu A^{\mu} = 0$, $n^2 = 0$. This is necessary for the
process-independent derivation of $\D$ because in such gauges
%
%
non-vanishing contributions to the fragmentation function are
produced only by diagrams where collinear radiation is both
emitted and absorbed by the same parton. In this paper we are
concerned with the case of collinear decay of a gluon. Examples of
relevant diagrams are shown on Fig.(1).

For the derivation of the fragmentation function we need to
consider only the spin-averaged case
\footnote{In general, in the case of decaying gluon the matrix
elements do not factorize due to the presence of non-trivial spin
correlations \cite{CG,CampGlover}. }
. Then, in the collinear limit Eq.(\ref{colllim}), matrix element
factorization holds in the following form:
\begin{eqnarray}
&& \vert M^{(n+2)}(k_1,\dots,k_{n-1},q_1,q_2,q_3)\vert^2 = \vert
M^{(n)}(k_1,\dots,k_{n-1},p)\vert^2~ W(n,\widehat{p},q_2,q_3)
+{\mathcal{O}}(\kappa). \label{MfactW}
\end{eqnarray}
Here $|M^{(n)}(....,p)|^2$ is the squared matrix element for
producing $n$ on-shell particles with non-exceptional momenta. All
the contribution to the fragmentation function from the collinear
splitting of a gluon is contained in the function $W$. The form of
that function in presence of massless fermions is given in
\cite{CG}:
\begin{equation}
W = -{1\over 2(1-\e)}g^{\mu\nu} V_{\mu\nu}. \label{W}
\end{equation}

The factor $V_{\mu\nu}$ represents the squared, color averaged sum
of diagrams contributing to the decay of an off-shell gluon with
momentum given by Eq.(\ref{phatsum}).

It is easy to generalize this result to the case when the final
state quark and antiquark have non-zero mass $m$. Repeating the
considerations in \cite{CG} one notices that the inclusion of
non-zero masses does not introduce new tensor structure in the
tensor decomposition of the function $V_{\mu\nu}$. Then, since the
limit $m\to 0$ is regular, the only effect of the inclusion of the
mass $m$ is to alter the corresponding scalar form-factors in that
tensor decomposition. Finally, we recall that in the collinear
limit the mass $m$ is considered to be of the order of the
transverse momenta of the collinear particles. Therefore, we
conclude that the inclusion of masses does not change the
dimensionality of the scalar form-factors or their scaling in the
collinear limit, i.e. in the presence of non-zero quark masses the
function $W$ is also given by Eq.(\ref{W}).

Combining Eqs.(\ref{fac},\ref{z},\ref{PSfact},\ref{MfactW}) we
find that the function $\widetilde{D}_{g/\Q}(z)$ appearing in
Eq.(\ref{DHresult}) can be written as:
\begin{eqnarray}
\widetilde{D}_{g/\Q}(z) = {1\over z} \int~W~ [dq_2;m][dq_3;0]
\delta\left(1 -z - {(nq_2)\over (pn)} -{(nq_3)\over (pn) } \right)
. \label{Dresult}
\end{eqnarray}

Similar expression holds for the one-loop virtual contribution to
that function. In this case one should replace in
Eq.(\ref{Dresult}) the phase-space factor $[dq_3;0]$ with the
corresponding phase-space for the virtual integration and omit the
term with $q_3$ in the energy-conservation delta-function. In both
cases the function $W$ is evaluated from Eq.(\ref{W}) and the
tensor $V_{\mu\nu}$ is constructed from the appropriate diagrams.
Two examples of such diagrams are shown on Fig.(1).
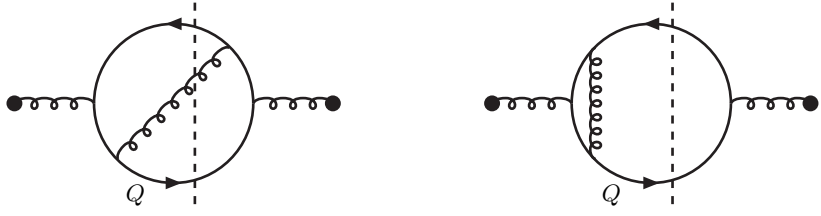
\begin{figure}
\begin{center}
\begin{picture}(200,40)(0,-20)
\SetWidth{1}
\Vertex(-50,10){3}
\Gluon(-50,10)(-20,10){2}{3}
\ArrowArc(10,10)(30,0,180)
\ArrowArc(10,10)(30,180,0)
\Gluon(40,10)(70,10){2}{3}
\DashLine(18,-28)(18,48){3}
\Gluon(-11,-11)(31,31){2}{7}
\Vertex(70,10){3}
\Vertex(130,10){3}
\Gluon(130,10)(160,10){2}{3}
\ArrowArc(190,10)(30,0,180)
\ArrowArc(190,10)(30,180,0)
\Gluon(220,10)(250,10){2}{3}
\DashLine(198,-28)(198,48){3}
\Gluon(169,-11)(169,31){2}{7}
\put(-8,-27){$Q$}
\put(172,-27){$Q$}
\Vertex(250,10){3}
\end{picture}
\end{center}
\caption{Examples of diagrams for the gluon decay process
$g \to Q + X$ at ${\cal O}(\alpha_s^2)$ that contribute to the
function $W$. The dashed vertical line indicates the intermediate
state that has to be considered.}
\end{figure}

\section{Results}

The first step in the evaluation of $\widetilde{D}_{g/\Q}(z)$ is
to obtain the splitting function $W$ in Eq.(\ref{W}). Our
calculation generalizes the tree-level massless results in
\cite{CG,CampGlover}. The next step is the integration over the
corresponding real and/or virtual momenta. To perform both the
phase-space and loop integrals, we follow the methods described in
\cite{AM,ADM,ADMP}: we first present all non-trivial phase-space
integrals as loop integrals and then use standard multiloop
methods \cite{tkachov}, such as integration-by-parts and
recurrence relations, to reduce all the phase-space and loop
integrals that have to be evaluated to a set of 18 master
integrals. For the reduction to master integrals we use the
algorithm \cite{Laporta} implemented in \cite{babis}. All
algebraic manipulations have been performed using Maple
\cite{MAPLE} and Form \cite{FORM}.

The coupling renormalization requires the result for the function
$\widetilde{D}_{g/\Q}(z)$ at order $\as$ including terms of order
${\cal O}(\e)$. The corresponding result can be found in
\cite{MM}. Field renormalization is performed by calculating
one-loop self-energy insertions in the external quark legs. The
use of light-cone gauge does not lead to appearance of
singularities that are not regulated by dimensional
regularization. As a last step we perform the collinear
renormalization according to Eqs.(\ref{Gamma}) and
(\ref{DHresult}).

We decompose the result for $d^{(2)}_{g}(z,\mu_0/m)$ in terms of
the color factors:
\begin{eqnarray}
d^{(2)}_{g}\left(z, {\mu_0\over m}  \right) &=& C_FT_R
F^{(C_FT_R)}_ {g} +C_AT_R F^{(C_AT_R)}_ {g} + T_R^2 F^{(T_R^2)}_
{g} + T_R^2n_l F^{(T_R^2n_l)}_ {g},
\label{d2Qresult}
\end{eqnarray}
where $n_l = n_f -1 $ is the number of massless flavors. The
functions $F_g$ read:
\begin{widetext}
\begin{eqnarray}
&& F^{(C_FT_R)}_ {g}  =
\Bigg\{-{4z^2-2z+1\over 2}\ln(z)+z-{1\over 4}+(2z^2-2z+1)\ln(1-z)\Bigg\}L^2
+\Bigg\{2(4z^2-6z+3)\Li_2(z)\nonumber \\
&& -{4z^2-2z+1\over 2}\ln^2(z)+6(2z^2-2z+1)\ln(1-z)\ln(z)
-3(2z^2-2z+1)\ln^2(1-z) +{8z-3\over 2}\ln(z)\nonumber\\
&&-(4z+1)\ln(1-z)-{10z^2-13z+8\over 2} +{2\pi^2z^2\over 3}\Bigg\}L
-16(2z^2-2z+1)\left(\Li_3\left({2z-1\over
z}\right)+\Li_3\left({2z-1\over z-1}\right) \right)\nonumber \\
&&-2(8z^2-10z+5)\Li_3(z)-16(2z^2-2z+1)\Li_3(1-z) +\Bigg\{
-8(2z^2-2z+1)\ln(z)+8(2z^2-2z+1)\ln(1-z)\nonumber\\
&& +{8(1-2z)^3\over 3}\Bigg\}\Li_2\left({2z-1\over z}\right)
+\left( -12(2z^2-2z+1)\ln(1-z)
+(8z^2-10z+5)\ln(z)+{32z^3-24z^2+12z-1\over 6}\right)\nonumber \\
&&\times\Li_2(z)-{36z^2-34z+17\over 12}\ln^3(z)
+{5(2z^2-2z+1)\over 2}\ln^3(1-z)
-{29(2z^2-2z+1)\over 2}\ln(z)\ln^2(1-z)\nonumber \\
&&+{11(2z^2-2z+1)\over 2}\ln^2(z)\ln(1-z)
-{256z^3-372z^2+156z-29\over 24}\ln^2(z) -{32z^3-46z^2+22z-5\over
4}\ln^2(1-z)\nonumber \\
&&+{64z^3-87z^2+33z+5\over 3}\ln(z)\ln(1-z)+
\left( {96z^3-196z^2+255z-48\over
12} -(2z^2-2z+1){\pi^2\over 6}\right)\ln(z)\nonumber \\
&&-\left( {24z^3-65z^2+79z-26\over 3} - (2z^2-2z+1){11\pi^2\over
6}\right)\ln(1-z) -2\pi^2(2z^2-2z+1)\big|\ln(z)-\ln(1-z)\big|\nonumber \\
&&+{2\pi^2\over 3}\big| 1-2z\big|(1-2z)^2
+(32z^3-54z^2+6z+19){\pi^2\over 36}+(56z^2-60z+30)\zeta(3)-{
152z^2-173z+69\over 12} ~~ ,
\end{eqnarray}
\end{widetext}
\vskip 1mm
\begin{widetext}
\begin{eqnarray}
&& F^{(C_AT_R)}_ {g} =
\Bigg\{ (4z+1)\ln(z)+(2z^2-2z+1)\ln(1-z) -{9z^2-2z-14\over 6}+
{2\over 3z}\Bigg\}L^2 +\Bigg\{-8(2z^2-2z+1)\Li_2(z)\nonumber\\
&&-2(2z^2+2z+1)\Li_2(-z)+(6z+1)\ln^2(z)
-4(2z^2-2z+1)\ln(z)\ln(1-z)-2(2z^2+2z+1)\ln(z)\ln(1+z)\nonumber\\
&& +(2z^2-2z+1)\ln^2(1-z) -{2(2z^2+17z+2)\over
3}\ln(z)+{10z^2-10z+11\over 3}\ln(1-z) -{2\pi^2 z\over 3}
+{178z^2-95z+13\over 9} \nonumber \\
&& -{20\over 9z}\Bigg\}L
+2(2z^2+2z+1)\left[S_{1,2}(-z)-S_{1,2}(z^2)
\right] +8(2z^2-2z+1)\left(\Li_3\left({2z-1\over
z}\right)+\Li_3\left({2z-1\over z-1}\right)\right)
\nonumber\\
&&+(6z^2+18z+7)\Li_3(z)+2(2z^2-10z+1)\Li_3(1-z)+(2z^2+2z+1)\Li_3(-z)
+\Bigg\{ 4(2z^2-2z+1)\ln(z) \nonumber\\
&&-4(2z^2-2z+1)\ln(1-z) -{4(1-2z)^3\over 3}\Bigg\}\Li_2
\left({2z-1\over z}\right)
+\Bigg\{-(6z^2+2z+5)\ln(z)+2(2z^2-10z+1)\ln(1-z)\nonumber \\
&&-4(2z^2+2z+1)\ln(1+z) -{8z^3+4z^2-33z-7\over 3}+{4\over 3z}
\Bigg\}\Li_2(z) +\Bigg\{-(2z^2+2z+1)\ln(z)-2(2z^2+2z+1)
\nonumber \\
&&\times\ln(1+z)+2z(1+z)\Bigg\}\Li_2(-z)+{8z^2+2z+5\over
6}\ln^3(z) -{7(2z^2-2z+1)\over 6}\ln^3(1-z)
+(10z^2-18z+5)\nonumber \\
&&\times\ln(z)\ln^2(1-z)-3(2z^2-2z+1)\ln^2(z)\ln(1-z)
-{2z^2+2z+1\over 2}\ln^2(z)\ln(1+z) -(2z^2+2z+1)\nonumber \\
&&\times\ln(z)\ln^2(1+z)+{32z^3-50z^2+25z-6\over 6}\ln^2(z)
-\left({64z^3-44z^2+10z-15\over 6}-{4\over
3z}\right)\ln(z)\ln(1-z)\nonumber \\
&&+2z(1+z)\ln(z)\ln(1+z)+{48z^3-44z^2-16z+23\over 12}\ln^2(1-z)
-\left({24z^3-624z^2-475z-181\over 18} \right.\nonumber \\
&&\left. +(2z^2+6z+1){\pi^2\over 6}\right)\ln(z) +\left(
(22z-6z^2-3){\pi^2\over 6} +{12z^3+76z^2-28z-22\over
9}\right)\ln(1-z) \nonumber \\
&&+{\pi^2\over 2}(2z^2+2z+1)\ln(1+z)+\pi^2(2z^2-2z+1)\big|
\ln(z)-\ln(1-z)\big| -{\pi^2\over 3}\big| 1-2z\big|(1-2z)^2
\nonumber \\
&&-(16z^3+24z^2+12z+29){\pi^2\over 36}-2(10z^2-3z+7)\zeta(3)
-{2568z^2-1583z-1331\over 54}+{56\over 27z} ~~ ,
\end{eqnarray}
\end{widetext}

\vskip 2mm

\begin{widetext}
\begin{eqnarray}
&& F^{(T_R^2)}_ {g} = -{2(2z^2-2z+1)\over 3}L^2
+\left\{-{4(2z^2-2z+1)\over 3}( \ln(z)+\ln(1-z) )-
{4(4z^2-4z+5)\over 9}\right\}L \nonumber \\
&&-{2z^2-2z+1 \over 3}(\ln^2(z)+\ln^2(1-z) )
+2(2z^2-2z+1)\ln(z)\ln(1-z)\nonumber \\
&&-{128z^5-320z^4+160z^3+40z^2-160z+50\over 45}\ln(z)
+{128z^5-320z^4+160z^3+120z^2-240z+102\over 45}
\ln(1-z)\nonumber \\
&&+(2z^2-2z+1){\pi^2\over 3} +{384z^4-768z^3+568z^2-184z-280\over
135} ~~ ,
\end{eqnarray}
\end{widetext}

\vskip 2mm

\begin{widetext}
\begin{eqnarray}
&& F^{(T_R^2n_l)}_ {g} = -{2(2z^2-2z+1)\over
3}L^2+\left\{-{4(2z^2-2z+1)\over
3}(\ln(z)+\ln(1-z))-{4(4z^2-4z+5)\over 9}\right\}L -{2z^2-2z+1\over 3}
\nonumber \\
&&\times\left[\ln(z)+\ln(1-z) \right]^2 -{2(4z^2-4z+5)\over
9}\left[\ln(z)+\ln(1-z)\right] +(2z^2-2z+1){\pi^2\over
3}-{88z^2-88z+56\over 27} ~~ ,
\end{eqnarray}
\end{widetext}
where $L=\ln(\mu_0^2/m^2)$.

Eq.(\ref{d2Qresult}) is the main result of this paper.

We next comment on the properties of Eq.(\ref{d2Qresult}). It
contains polylogarithmic functions up to rank three. These
functions are defined through:
\begin{equation}
\Li_n(z)=\int_0^z{\Li_{n-1}(x)\over x} dx , ~~ \Li_1(z)=-\ln(1-z)
, ~~~S_{1,2}(z)={1\over 2}\int_0^z{\ln^2(1-x)\over x} dx .
\nonumber
\end{equation}
The mass-enhanced terms proportional to
$\ln^k(\mu_0^2/m^2)~,~k=1,2$ are of the general form discussed in
\cite{MM}. The behavior of Eq.(\ref{d2Qresult}) in the limits
$z\to 0$ and $z\to 1$ is as follows: in the limit $z\to 1$ the
gluon initiated initial condition is only as singular as
$\sim\ln^3(1-z)$ and therefore suppressed with an inverse power of
$N$ for large values of the Mellin variable. For small $z$, the
function $d^{(2)}_{g}$ exhibits $\sim C_AT_R/z$ pole and, in
addition, has less singular terms $\sim\ln^n(z)~,n\leq 3$.

There are two more terms in $d^{(2)}_{g}$ that deserve comment.
They contain the functions $\vert \ln(z)-\ln(1-z)\vert$ and $\vert
1-2z\vert$. Those terms are proportional to the color factor
$(C_F-C_A/2)T_R$ and originate from the virtual diagram
representing the quark-gluon vertex correction shown on Fig.(1).
To interpret these terms we first rewrite them as $2{\rm arctanh}(
\sqrt{(1-2z)^2})$ and $\sqrt{(1-2z)^2}$ respectively. We see that
in the point $z=1/2$ these functions are not smooth.

At the point $z=1/2$ the virtual diagram in question represents a
configuration where a quark-antiquark pair is created and the both
final-state particles have the same energy. Since their relative
transverse momentum is small, that configuration represents
effectively a threshold production of a heavy pair \cite{Kirill}.
It is well known (see \cite{Hoang} for example) that the one-loop
vertex correction for heavy pair creation close to threshold
contains powers $\beta^n~,~n\geq -1$ of the velocity:
\begin{equation}
\beta = \sqrt{1-{4m^2\over Q^2}} ,
\nonumber
\end{equation}
where $m$ is the mass of the quark and $Q$ the invariant mass of
the pair.

One can easily verify that in our collinear kinematics the
analogue of $\beta$ takes the form:
\begin{equation}
\sqrt{1-{4z(1-z)\over 1+\vert q_{\bot}^2\vert /m^2}} , \nonumber
\end{equation}
and for vanishing relative transverse momentum $q_{\bot}$ between
the two final-state particles the threshold condition is indeed
$z=1/2$. With an explicit calculation one can also check that
terms from the virtual diagram that are of the form
$\gamma^\mu\beta^{2n-1} ,~n \geq 0$, indeed lead to contributions
to $d^{(2)}_{g}$ that are proportional to the functions ${\rm
arctanh}(\sqrt{(1-2z)^2})$ and $\sqrt{(1-2z)^2}$.

\section{Conclusions}

In this paper we compute the gluon initiated contribution to the
initial condition for the perturbative fragmentation function of a
heavy quark through order ${\mathcal{O}}(\as^2)$ in the $\overline
{\rm MS}$ scheme. To derive this result we follow the general
approach developed in \cite{MM}. The result has the following
properties: the terms with mass logarithms are of the general form
predicted by the DGLAP equation. We find that the gluon initiated
contribution is suppressed with an inverse power of $N$ in the
limit $N\to\infty$, where $N$ is the conjugated to $z$ Mellin
variable. The result also exhibits a non-smooth behavior at the
mid-point $z=1/2$ which we interpret as due to Coulomb interaction
at threshold. Unlike the order ${\cal O}(\alpha_s)$ result, the
gluon initiated initial condition at order ${\cal O}(\alpha_s^2)$
does not vanish when $m=\mu_0$. Our calculation shows the presence
of non-trivial constant term.

Combined with the results for the fermion-initiated contributions
derived in \cite{MM} this result completes the calculation of the
initial condition for the perturbative fragmentation function
through order ${\mathcal{O}}(\as^2)$. After the three-loop
time-like splitting functions become available, our results will
permit the resummation with next-to-next-to-leading logarithmic
(NNLL) accuracy of the large quasi-collinear logs $\sim
\ln^k(m^2)$ that appear in heavy quark differential distributions,
i.e. will extend the Perturbative Fragmentation Function formalism
\cite{MN} to the NNLL level.

Our result has large number of potential applications due to the
process independence of the PFF method. Of central importance will
be its application to $e^+e^-$ in order to extract with NNLL
accuracy the non-perturbative fragmentation components. With such
result at hand, one can make high precision predictions for heavy
flavor energy and transverse momentum distributions at hadron
colliders, top-decay, DIS, etc. Combined with the previously
calculated fermion initiated components, this result can also be
used to evaluate spectra of heavy quarks at NNLO in any process
and up to power corrections $(m/Q)^n$, from purely massless
calculations.

{\bf Acknowledgments.} I would like to thank Kirill Melnikov for
numerous useful discussions, for careful reading of the manuscript
and for his help with solving the IBP identities using the program
\cite{babis}. This research is supported by the DOE under contract
DE-FG03-94ER-40833 and by the start up funds of the University of
Hawaii.

\end{document}